\documentclass[a4paper,conference]{IEEEtran}
\usepackage{cite}
\usepackage{amsmath,amssymb,amsfonts}
\usepackage{algorithmic}
\usepackage{graphicx}
\usepackage{textcomp}
\usepackage{xcolor}
\usepackage{float}
\usepackage{booktabs}
\usepackage{url}
\usepackage{placeins}
\usepackage{csquotes}
\usepackage{array}
\usepackage{longtable}
\usepackage{cuted}

\begin{document}

\title{``I Wrote, I Paused, I Rewrote'': A Mixed-Methods Study of Learner Perceptions of Process-Aware LLM Feedback\\}

\author{\IEEEauthorblockN{Samra Zafar}
\IEEEauthorblockA{samrazafar003@gmail.com}
\and
\IEEEauthorblockN{Shifa Yousaf}
\IEEEauthorblockA{shifayousaf4@gmail.com}
}


\maketitle


\begin{abstract}
Large language models increasingly provide formative writing feedback,
but most systems evaluate only the submitted text. Process-aware systems
can additionally incorporate intermediate drafts and interaction records,
yet access to these traces does not establish whether learners perceive
the resulting feedback as useful, fair, or aligned with their own
revision experience. We conducted a convergent mixed-method post-task
evaluation with 43 undergraduate English-language learners. Participants
completed a timed essay, received LLM-generated feedback based on the
final essay, intermediate snapshots, and keystroke-derived records, and
then answered six five-point Likert items and four open-ended questions.

Ratings were concentrated above the neutral midpoint across all six
items. Between 74.4\% and 95.3\% of participants selected ratings of 4
or 5, and all six one-sided exact sign tests remained significant after
Holm correction. Perceived correspondence between the revision analysis
and the participant's reported thought process was positively associated
with overall perceived usefulness, Spearman's $\rho_s=.327$, 95\%
bootstrap CI $[.009,.609]$, $p=.032$, $n=43$.

Qualitative responses were analysed through an LLM-assisted codebook
development procedure followed by independent human coding of 14 binary
codes. Participants frequently described the product feedback and
revision rationale as accurate, while a smaller set reported
misinterpretation of surface corrections, time constraints, or rhetorical
intention. The findings distinguish perceived detection of revision
activity from perceived interpretation of why revision occurred.
\end{abstract}


\section{Introduction}
In schools, Large Language Models (LLMs) are increasingly used to
provide feedback on student writing. They’re easy to get, understood quickly and usually helpful, though a problem can occur. Most software today responds only to the final draft of a text and
gives feedback that is static and general. In turn, this lets
educators forget that students didn’t choose their issues; the
challenges affected their progress along the way. The task of
writing means constantly stop- ping, starting again, fixing up the
work and dealing with moments you’re not sure about. If
feedback doesn’t focus on the student’s past experience, it may
just provide general tips—instead of helping them grow in
learning \cite{meyer2025individualdifferences}.

Much of the research community now supports the idea that good
feedback covers the whole process, not just the final outcome. It
should represent how students actually form their arguments. As
illustrated by Schiller et al. \cite{schiller2024processdata}, static systems
don’t see that there might be varying student engagement
depending on the stage of writing—being it the initial planning
or mid-writing stage. Even though using LLMs increases
interest in the subject \cite{meyer2024llmfeedback}, their lack of thought processing capabilities decreases what they can offer in education.
This is important because taking breaks, preparing work and
changing your plan are big parts of becoming a strong writer \cite{flower1981cognitive,leijten2013keystroke}. To help with understanding
and supporting those skills, a few researchers are using things
like records of keystrokes and frequently captured drafts by
students. They make it possible to observe students’ writing as it
happens, seeing what they write and how they approach it. With
ScholaWrite \cite{wang2025scholawrite}, more details are collected at each
moment to display how students prepare, change and think about
their written work. Another study done by Meyer and Jansen \cite{meyer2025individualdifferences} found that both early performance and motivation affected the way students react to feedback, implying that context-specific, personalized systems could be valuable.

Process-aware writing systems attempt to address this limitation by
incorporating intermediate text states and interaction records into
feedback generation. Keystroke-derived records and periodic snapshots
can make observable revision activity available to an LLM. Nevertheless,
access to process records creates two distinct interpretive tasks. A
system may detect that a pause, deletion, or reorganization occurred,
but it must additionally infer why that event occurred. A backspace
sequence may reflect a typing error, lexical refinement, conceptual
uncertainty, or time pressure. Consequently, process detection does not
necessarily imply accurate interpretation of learner intention.

This distinction is important for learner-facing educational systems.
Even feedback that correctly identifies observable activity may be
experienced as unfair or inaccurate when it assigns an unsupported
reason to that activity. Conversely, learners may value process-aware
feedback when it identifies difficulties that they recognize in their
own writing experience. System-side analysis alone therefore cannot
establish whether learners perceive process-aware feedback as relevant,
interpretable, or useful.

We conducted a convergent mixed-method post-task evaluation of an
LLM-based writing-feedback system that used final essays, three-minute
snapshots, and keystroke-derived revision records. Participants evaluated
the resulting feedback through six Likert items and four open-ended
questions. The quantitative strand examined learner ratings of feedback
relevance, process correspondence, usefulness, and fairness. The
qualitative strand examined where learners perceived correspondence,
misinterpretation, or omission in the system's treatment of their
writing process.

\textbf{RQ1:} How do learners rate the perceived relevance, process correspondence, usefulness, and prompt-related fairness of process-aware LLM writing feedback?

\textbf{RQ2:} How do learners describe correspondence and divergence between process-aware feedback and their revision experience, and what improvements do they request?

This study contributes: (1) a learner-centered evaluation of
process-aware LLM feedback; (2) a quantitative analysis of perceived
feedback relevance and process correspondence; (3) an independently human-coded thematic content analysis with code-level reliability assessment; and (4) an empirical
distinction between detecting revision activity and interpreting the
reason for that activity.

\section{Related Work}
Recent research has examined whether LLM-generated writing feedback can achieve the specificity and instructional value expected from human feedback. Steiss et al. compared feedback produced by ChatGPT with feedback written by human reviewers and identified both promising quality characteristics and areas where pedagogical judgment remained important \cite{steiss2024humanChatGPT}. Similarly, Dai et al. evaluated GPT-generated feedback in higher-education writing tasks and reported differences among GPT models and instructors in readability, feedback components, consistency, and assessment reliability \cite{dai2024automaticfeedback}. These studies demonstrate the potential of LLMs for scalable assessment support, but their evaluations primarily consider the submitted response rather than the sequence of revisions that produced it.

Other work has focused on improving the instructional effectiveness of generated feedback. Nair et al. proposed a framework that optimizes writing feedback according to its expected effect on subsequent revision, using language-model-simulated student revisions \cite{nair2024closingloop}. This approach emphasizes that feedback quality should be evaluated through its capacity to support revision, rather than fluency alone. However, the simulated revision process does not provide direct evidence of how an individual learner actually developed a particular essay.

Writing-process research provides methods for collecting such evidence. Leijten and Van Waes showed how keystroke logging can reconstruct text production and support analysis of revisions, pauses, and writing strategies \cite{leijten2013keystroke}. They also cautioned that fine-grained interaction records cannot, by themselves, establish the cognitive reasons behind an observed action. Building on these strands, the present study incorporates timestamped writing traces into learner-facing LLM feedback and examines how learners perceive its usefulness and alignment with their revision experience, including concerns about fairness and contextual interpretation.

\section{Study Design and Methods}
\subsection{Study Design}

We used a convergent mixed-method post-task design. Participants first
completed a timed writing task and reviewed process-aware LLM feedback.
Quantitative and qualitative responses were then collected through the
same survey. The two strands were analysed separately and subsequently
integrated to examine convergence and divergence between item-level
ratings and open-ended learner accounts.

\subsection{Participants and Procedure}

The analysed sample comprised 43 undergraduate students who were
non-native speakers of English and studied in an English-medium
university context. Participants were recruited through academic
channels and voluntary sign-up. Each participant completed a
20-minute essay-writing task and received feedback generated from the
submitted essay and recorded writing-process data. All 43 participants
subsequently completed the post-task survey and were included in the
analysis.

Before participation, learners were informed that the interface would
record intermediate text states and revision-related activity.
Participation was voluntary, responses were anonymized, and the
feedback was not used for course grading.

\subsection{Writing Interface and Feedback Generation}

The study used a browser-based writing interface with a JavaScript
frontend and Python backend. Participants received one randomly selected
prompt from a set of 15 argumentative, reflective, and analytical
topics and completed a 20-minute essay-writing task.

The interface recorded timestamped intermediate text states associated
with backspace activity. Consecutive backspaces were grouped until no
backspace had occurred for three seconds, reducing redundant records.
Textbox snapshots were also captured at three-minute intervals. At
submission, the final essay, intermediate snapshots, and timestamped
revision records were provided to the Gemini API.

The generated feedback contained two components. The first addressed
thesis and argumentation, language use, prompt relevance, and
organization. The second described revision-related patterns, including
pauses, deletions, reformulation, and changes across snapshots. Statements
about learner intention or cognition were model-generated interpretations
rather than direct measurements. Following are a few examples of the LLM revision feedbacks,
showing the time and context awareness of the LLM:

\textit{“Snapshots captured between the 3–6 minute mark
reveal the gradual development of main arguments
regarding part-time jobs. Nonetheless, a prolonged pause
followed by substantial rewriting near the conclusion
suggests difficulty in articulating a strong closing statement.”}

\textit{“A marked pause around the 15-minute point precedes the
addition of content related to positive outcomes, indicating a
potential strategic shift in the student’s argumentative
framing.”}

\subsection{Post-Task Survey Instrument}

After reviewing the generated feedback, participants completed a
post-task survey containing six five-point Likert items and four
open-ended questions. The Likert items assessed: (1) perceived accuracy
of feedback on thesis and argument structure; (2) perceived relevance
and helpfulness of language and grammar feedback; (3) whether the system
identified areas of revision or hesitation; (4) whether the revision
analysis corresponded with the participant's perceived thought process;
(5) overall perceived usefulness for improving writing; and (6) the
perceived fairness and insightfulness of comments on engagement with the
prompt. Responses ranged from 1 (\emph{strongly disagree}) to 5
(\emph{strongly agree}), with 3 representing a neutral response.

The open-ended questions asked participants to identify the most accurate
or helpful aspect of the feedback, describe any misunderstanding of
their thinking or intention, explain whether the system understood why
they revised particular parts of the essay, and recommend improvements
for future feedback. The items were developed specifically for this
post-task evaluation and were therefore analysed individually rather
than combined into a single psychometric scale.

\subsection{Quantitative Analysis}

Likert items were treated as ordinal variables. For each item, we
calculated the median, interquartile range (IQR), mean, standard
deviation, and the number and percentage of participants selecting
ratings of 4 or 5. Medians, IQRs, and response distributions were treated
as the primary descriptive summaries.

One-sided exact sign tests examined the directional hypothesis that,
among non-neutral responses, ratings above the neutral midpoint of 3
would occur more frequently than ratings below 3. Ratings above 3 were
classified as positive, ratings below 3 as negative, and ratings equal
to 3 as ties. Ties were excluded from the effective sample size.
Holm's sequential procedure adjusted the six item-level $p$-values to
control the family-wise error rate.

As an exploratory analysis, Spearman's rank correlation examined the
association between perceived correspondence of the revision analysis
with the participant's reported thought process (Item 4) and overall
perceived usefulness for improving writing (Item 5). Only participants
with valid responses to both items were included. The coefficient was
calculated in Python using SciPy. A 95\% percentile bootstrap confidence
interval was estimated from 10,000 paired participant-level resamples
using a fixed random seed. The correlation is interpreted as an
association between perceptions rather than as evidence of causation.

Descriptive statistics, exact sign tests, and Holm adjustments were
calculated in Google Sheets. Statistical significance was evaluated at
$\alpha=.05$.

\subsection{Qualitative Analysis}

Open-ended responses were analysed using inductive--deductive thematic
content analysis. The authors first examined a development subset of
10 participant responses. ChatGPT was used during this stage as a
code-generation and code-organization aid. The authors then reviewed,
merged, renamed, and operationally defined the proposed categories.
This process produced a frozen codebook containing 14 binary codes.

The codes captured perceived product-feedback accuracy, recognition of
process activity, recognition of revision rationale, partial alignment,
process misinterpretation, conflation of surface and substantive
revision, mismatch with time or word-count constraints, overemphasis on
corrected errors, intention or style mismatch, requests for actionable
guidance, readability, rubric transparency, interface improvements,
and explicit reports of no misunderstanding.

Two research assistants independently applied all 14 codes to the
complete set of 43 participant responses. A code was marked 1 when it
was present anywhere in the participant's four open-ended answers and
0 when it was absent. Codes were non-mutually exclusive. Inter-rater
reliability was calculated before adjudication using raw percentage
agreement and Cohen's $\kappa$ for each binary code.

Across the 14 codes, raw agreement ranged from 88.4\% to 100\% and
Cohen's $\kappa$ ranged from .449 to 1.00, with a median of .774.
Across 602 participant--code decisions, the coders agreed on 573
(95.2\%). Disagreements were subsequently adjudicated against the frozen
codebook and raw response text. The final adjudicated dataset was used
to calculate code prevalence and organize codes into higher-level
themes. ChatGPT generated codebook-based recommendations for disagreement cases. An author independently reviewed the raw response, disputed code, and inclusion/exclusion criteria and made the final adjudication decision.

\subsection{Mixed-Methods Integration}

The quantitative and qualitative strands were initially analysed
separately. They were subsequently integrated through a joint display
linking each survey construct with corresponding qualitative themes.
Evidence was classified as convergent when the Likert distribution and
open-ended accounts supported a similar interpretation, complementary
when the qualitative data added an issue not directly measured by the
Likert items, and divergent when generally positive ratings coexisted
with reports of misinterpretation or contextual mismatch. Integration
focused particularly on the distinction between detecting that revision
activity occurred and interpreting why the learner revised.

\section{Results}

\subsection{Participant Flow and Response Completeness}

A total of 52 participants completed the essay-writing task and
received process-aware feedback. Of these, 43 submitted the
post-task survey, corresponding to a response rate of 82.69\%.
No record was excluded due to duplication or incompleteness. All six Likert items had 43 valid responses. All 43 participants provided valid responses to Items 4 and 5 and were included in the correlation analysis. Open-ended responses from all 43 participants were included in the qualitative dataset, although individual questions contained brief or non-substantive responses such as “N/A” or “I don’t know.”
\begin{table*}[t]
\centering
\caption{Learner ratings of process-aware LLM feedback.}
\label{tab:likert_results}
\begin{tabular}{lrrrrrrrr}
\toprule
Item & $N$ & Median & IQR & Ratings 4--5, $n$ (\%) &
Positive & Negative & Ties \\
\midrule
Thesis and argument accuracy
  & 43 & 4 & 1 & 76.6 & 33 & 1 & 9  \\
Language and grammar relevance
  & 43 & 4 & 1 & 88.3 & 38 & 0 & 5  \\
Revision/hesitation detection
  & 43 & 4 & 1 & 95.3 & 41 & 0 & 2  \\
Perceived process correspondence
  & 43 & 4 & 1.5 & 74.4 & 32 & 2 & 9  \\
Overall usefulness
  & 43 & 4 & 1 & 86.0 & 37 & 1 & 5  \\
Prompt-engagement fairness
  & 43 & 4 & 1 & 88.3 & 38 & 0 & 5  \\
\bottomrule
\end{tabular}
\end{table*}

\subsection{RQ1: Learner Perceptions of Process-Aware Feedback}

Table~\ref{tab:likert_results} summarizes the six Likert items. All six items
had a median of 4, although the response distributions differed.
Revision and hesitation detection received the largest proportion of
ratings at 4 or 5, with 41 of 43 participants (95.3\%) selecting an
upper-scale response. Language and grammar relevance and
prompt-engagement fairness were each rated 4 or 5 by 38 participants
(88.4\%). Overall usefulness received upper-scale ratings from 37
participants (86.0\%).

Perceived process correspondence received the lowest proportion of
ratings at 4 or 5, although the distribution remained predominantly
positive: 32 participants (74.4\%) selected 4 or 5, nine selected the
neutral midpoint, and two selected a rating below neutral. Its IQR of
1.5 and standard deviation of 0.94 were also larger than those of the
other items. Descriptively, participants showed greater consensus that
the system detected revision activity than that it accurately reflected
their reported thought process.

The one-sided exact sign tests supported the directional hypothesis that
non-neutral ratings would occur more frequently above than below the
neutral midpoint for all six items. All six results remained significant
after Holm correction, with all adjusted $p<.001$. Because the sign tests were directional, their interpretation is limited
to the prespecified expectation of predominantly positive learner
ratings and should not be treated as a general test of any departure
from neutrality.

\begin{figure}[t]
    \centering
    \includegraphics[width=\columnwidth]{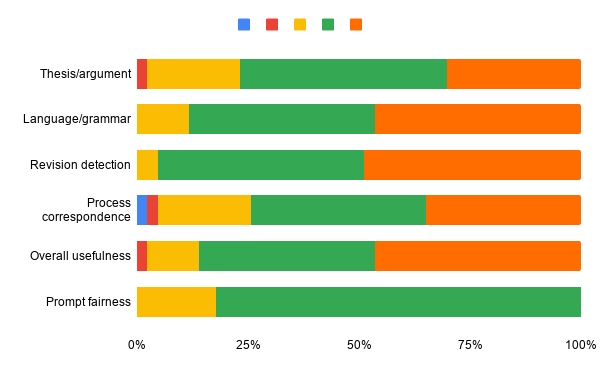}
    \caption{Distribution of participant responses across the six Likert
    items. Ratings ranged from 1 (\emph{strongly disagree}) to 5
    (\emph{strongly agree}).}
    \label{fig:likert_distribution}
\end{figure}

Figure~\ref{fig:likert_distribution} shows that responses were
concentrated in the upper categories across all six items, with the
strongest concentration for revision and hesitation detection.

\subsection{Association Between Process Correspondence and Usefulness}

Perceived correspondence between the revision analysis and the
participant's reported thought process was positively associated with
overall perceived usefulness, Spearman's $\rho_s=.327$, 95\% bootstrap
CI $[.009,.609]$, two-sided $p=.032$, $n=43$. Participants who reported
greater perceived correspondence therefore tended to report greater
usefulness. However, the association was modest, the confidence interval
was wide, and the cross-sectional post-task design does not support a
causal interpretation.

\subsection{Qualitative Coding Reliability}

Before adjudication, raw agreement ranged from 88.4\% to 100\% across
the 14 binary codes. Cohen's $\kappa$ ranged from .449 to 1.00, with a
median of .774. Nine codes achieved $\kappa\geq.75$, four ranged from
.642 to .656, and one code produced a lower value of .449. The code
with the lowest $\kappa$ had low positive prevalence and four disputed
positive decisions, so its value was interpreted alongside its 90.7\%
raw agreement. Overall, the coders agreed on 573 of 602 binary coding
decisions (95.2\%).
\begin{table*}[t]
\centering
\caption{Final code prevalence and pre-adjudication inter-rater reliability
for the qualitative analysis ($N=43$).}
\label{tab:qual_codes}
\begin{tabular}{lrrr}
\toprule
Code & Final $n$ (\%) & Raw agreement & Cohen's $\kappa$ \\
\midrule
Product feedback accurate
& 35 (81.4\%) & 100.0\% & 1.000 \\

Process activity recognized
& 20 (46.5\%) & 90.7\% & .809 \\

Revision reason recognized
& 34 (79.1\%) & 90.7\% & .758 \\

Partial process alignment
& 9 (20.9\%) & 93.0\% & .780 \\

Process misinterpretation
& 4 (9.3\%) & 95.3\% & .645 \\

Surface revision conflation
& 5 (11.6\%) & 97.7\% & .876 \\

Time/word-count expectations mismatch
& 6 (14.0\%) & 97.7\% & .896 \\

Corrected errors overemphasized
& 2 (4.7\%) & 97.7\% & .656 \\

Intention or style mismatch
& 3 (7.0\%) & 95.3\% & .642 \\

Actionable guidance requested
& 20 (46.5\%) & 88.4\% & .768 \\

Feedback readability requested
& 2 (4.7\%) & 97.7\% & .656 \\

Rubric transparency requested
& 1 (2.3\%) & 100.0\% & 1.000 \\

System speed or interface request
& 5 (11.6\%) & 90.7\% & .449 \\

No reported misunderstanding
& 26 (60.5\%) & 97.7\% & .951 \\
\bottomrule
\end{tabular}
\end{table*}

Final prevalence values were calculated from the adjudicated coding
dataset, whereas agreement and $\kappa$ values were calculated from the
two independent coding sheets before adjudication.
\subsection{RQ2: Learner Accounts of Correspondence and Misinterpretation}

\subsubsection{Perceived Recognition of Product and Process Evidence}

The most frequent qualitative pattern was perceived accuracy of the
feedback. Product-level feedback on thesis, organization, language, or
prompt engagement was coded for 35 of 43 participants (81.4\%).
Recognition of the reason for revision appeared in 34 responses
(79.1\%), while explicit recognition of observable process activity,
such as pauses, backspacing, revision timing, or repeated correction,
appeared in 20 responses (46.5\%). Twenty-six participants (60.5\%)
explicitly reported no misunderstanding.

Participants commonly described the system as identifying where they
hesitated, revised introductions or conclusions, refined word choice, or
struggled to formulate ideas. These accounts indicate that many learners
recognized both product-level observations and process-related
interpretations as corresponding to their experience. However, these
reports remain learner perceptions and should not be treated as
independent verification of cognitive accuracy.

\subsubsection{Detection Without Consistently Correct Interpretation}

A smaller but conceptually important group of participants distinguished
between detecting revision activity and interpreting its cause. Nine
participants (20.9\%) described partial process alignment, four (9.3\%)
reported a process misinterpretation, and five (11.6\%) described surface
corrections such as spelling, typing, or phrasing changes being treated
as deeper substantive revision.

For example, one participant reported that the system interpreted
backspacing as self-doubt, although the revisions resulted from fast
typing and spelling correction. Another agreed that the system detected
fewer revisions near the conclusion but rejected its interpretation that
this represented stronger focus, explaining that the reduced revision
resulted from running out of time. These cases illustrate that observable
process evidence can support feedback while still permitting unsupported
interpretations of intent.

\subsubsection{Contextual and Fairness Boundaries}

Six participants (14.0\%) reported that the feedback did not adequately
account for time or word-count constraints. Three (7.0\%) described a
mismatch with rhetorical intention or writing style, and two (4.7\%)
reported that already-corrected errors were still emphasized.

These responses suggest that process awareness alone does not guarantee
contextually calibrated feedback. Participants objected when rushed
writing was interpreted as lack of focus, when deliberate examples were
treated as digressions, or when corrected spelling errors continued to
influence evaluative comments. Such concerns were less frequent than
positive evaluations but identify important design risks in
process-aware feedback.

\subsubsection{Requests for More Actionable and Usable Feedback}

Twenty participants (46.5\%) requested more actionable guidance,
including specific examples, suggested replacements, highlighted text
spans, vocabulary alternatives, and clearer explanations of how to
improve weaknesses. Two participants (4.7\%) requested greater
readability, one (2.3\%) requested clearer rubric transparency, and five
(11.6\%) requested interface or system-level improvements.

The requests indicate that perceived relevance did not eliminate the
need for more concrete support. Participants frequently wanted the
system to move from identifying a weakness to demonstrating how a
specific sentence, phrase, or structural choice could be revised.

\subsection{Mixed-Methods Integration}

The two strands converged most strongly on revision detection and general
usefulness. Item~3 received the highest upper-scale response rate
(95.3\%), while the qualitative analysis identified numerous accounts
of accurate pause, revision, and hesitation recognition. Product-feedback
accuracy was likewise common in both the Likert ratings and open-ended
responses. The lower qualitative prevalence should not be interpreted as disagreement with the Likert result, because code prevalence reflects spontaneous mention in open-ended responses rather than endorsement of a directly presented item.

The findings were more mixed for process interpretation. Item~4 remained
positively rated, but it had the lowest upper-scale response rate
(74.4\%) and the greatest dispersion. The qualitative data explained
this variation by identifying partial alignment, surface-revision
conflation, time-pressure mismatch, and intention-related disagreement.
Thus, the combined evidence suggests that learners more consistently
recognized the system's ability to detect revision activity than its
ability to explain why that activity occurred.

Actionability was primarily a qualitative finding. Although overall
usefulness was rated positively, nearly half of participants requested
more specific and directly usable guidance. Positive usefulness ratings
therefore coexisted with clear expectations for more concrete,
text-linked recommendations.

\section{Discussion}

The findings indicate that learners generally perceived the
process-aware feedback as relevant and useful. All six Likert items were
rated above the neutral midpoint, and the largest consensus concerned
the system's identification of revision or hesitation. The qualitative
analysis similarly showed frequent recognition of product-level
feedback and revision rationale. These results support the acceptability
of process-aware feedback within this task context, but they do not show
that the system objectively understood learner cognition or improved
writing performance.

The main conceptual result is the distinction between revision detection
and revision interpretation. Learners could agree that the system
recognized a pause, backspace sequence, or reduced revision frequency
while disagreeing with the explanation assigned to that event. This
boundary was visible quantitatively: revision detection received more
consistently positive ratings than correspondence with the participant's
reported thought process. It was also visible qualitatively in accounts
of typing errors being interpreted as self-doubt, reduced revision being
interpreted as stronger focus, and time pressure being overlooked.

The positive association between perceived process correspondence and
overall usefulness suggests that alignment with the learner's own account
may matter for acceptance. However, the association was modest and its
confidence interval was wide. Both measures were collected from the
same survey after one exposure, so shared response tendencies and general
satisfaction may partly explain the relationship.

The qualitative findings also identify design requirements beyond
process detection. Nearly half of participants requested more actionable
guidance. Process-aware systems should therefore connect identified
events to specific text spans and concrete revision options rather than
merely describing the existence of hesitation or change. They should
also distinguish surface correction from substantive revision and allow
learners to contest or correct interpretations of their intention.

\section{Limitations}

This study used a single-group post-task design and did not compare
process-aware feedback with final-text-only or instructor-generated
feedback. The results therefore concern learner perceptions and do not
establish that process data improved feedback quality, writing
performance, or learning. The survey items were developed for this
study and were analysed individually rather than as a validated scale.
Perceived correspondence with thought process represents self-report,
not independent cognitive ground truth. The sample was drawn from one
English-medium undergraduate context, and several qualitative codes had
low prevalence, making their kappa estimates less stable.

\section{Ethical Considerations}

No formal institutional ethics-board review was obtained for this
low-stakes voluntary study. Before beginning the activity, participants
were shown an information notice describing the writing task, the
collection of intermediate text states and revision-related interaction
data, the research use of their responses, and the voluntary nature of
participation. Participants indicated their agreement by choosing to
proceed with the activity and could decline participation by leaving
before starting.
The activity was not connected to grading or other consequential
assessment. Records were anonymized using participant identifiers.
Because writing-process traces can reveal fine-grained patterns of
composition and revision, they were treated as potentially sensitive
data and interpreted cautiously. The participant notice is not included
in the anonymous submission because it contains information that may
reveal the study context.

\section{Conclusion}

This study evaluated learner perceptions of process-aware LLM writing
feedback through a convergent mixed-method post-task design. Participants
generally rated the feedback as relevant, useful, and capable of
identifying revision activity. Perceived correspondence with the
learner's reported thought process was positively associated with
overall usefulness.

The qualitative findings nevertheless identify an important boundary:
detecting that revision occurred is not equivalent to correctly
interpreting why it occurred. Learners valued recognition of pauses,
revisions, and writing difficulties, but some reported unsupported
interpretations of typing errors, time pressure, or rhetorical intention.
Future process-aware systems should therefore emphasize transparent,
contestable, and text-linked feedback rather than treating behavioural
traces as direct evidence of cognition.

\bibliographystyle{IEEEtran}
\bibliography{references}

\appendices
\section{Post-Task Survey Instrument}
\label{app:survey}
Participants completed a post-task survey after reviewing the
process-aware LLM feedback. The survey consisted of six five-point
Likert-scale items and four open-ended questions.

The Likert scale was defined as:
\begin{itemize}
    \item 1 = Strongly disagree
    \item 2 = Disagree
    \item 3 = Neutral
    \item 4 = Agree
    \item 5 = Strongly agree
\end{itemize}

\FloatBarrier
\subsection{Likert-Scale Items}
\begin{table}[h]
\centering
\caption{Post-task survey Likert-scale items.}
\label{tab:survey_items}
\begin{tabular}{lp{7.2cm}}
\toprule
Item & Survey Statement \\
\midrule
Q1 & The feedback accurately reflected my intended thesis and argument structure. \\
Q2 & The language and grammar feedback was relevant and helpful. \\
Q3 & The system identified areas where I revised or hesitated during writing. \\
Q4 & The analysis of my revision behavior reflected my actual thought process. \\
Q5 & Overall, the feedback was helpful for improving my writing skills. \\
Q6 & The comments about my engagement with the prompt were fair and insightful. \\
\bottomrule
\end{tabular}
\end{table}

\FloatBarrier
\subsection{Open-Ended Questions}

\begin{enumerate}
    \item Which part of the feedback felt most accurate or helpful?
    \item Was there any part that misunderstood your thinking or intention?
    \item Do you think the system understood why you revised certain
    parts of your essay? Please explain.
    \item What would you like the system to do better in future feedback?
\end{enumerate}

\FloatBarrier
\section{Qualitative Coding Codebook}
\label{app:codebook}
The qualitative analysis used an inductive--deductive thematic
content analysis approach. A frozen codebook containing 14 binary
codes was developed and applied independently by two research
assistants.

\begin{table}[h]
\centering
\caption{Mapping of qualitative codes to abbreviated labels.}
\label{tab:code_mapping}
\renewcommand{\arraystretch}{1.15}
\begin{tabular}{p{7.2cm}p{0.8cm}}
\toprule
\textbf{Original Code Name} & \textbf{Short Code} \\
\midrule
PRODUCT\_FEEDBACK\_ACCURATE & PFA \\
PROCESS\_ACTIVITY\_RECOGNIZED & PAR \\
REVISION\_REASON\_RECOGNIZED & RRR \\
PARTIAL\_PROCESS\_ALIGNMENT & PPA \\
PROCESS\_MISINTERPRETATION & PMI \\
SURFACE\_REVISION\_CONFLATION & SRC \\
TIME\_AND\_WORD\_COUNT\_EXPECTATIONS\_MISMATCH & TWC \\
CORRECTED\_ERRORS\_OVEREMPHASIZED & CEO \\
INTENTION\_OR\_STYLE\_MISMATCH & ISM \\
ACTIONABLE\_GUIDANCE\_REQUESTED & AGR \\
FEEDBACK\_READABILITY\_REQUESTED & FRR \\
RUBRIC\_TRANSPARENCY\_REQUESTED & RTR \\
SYSTEM\_SPEED\_OR\_INTERFACE\_REQUEST & SIR \\
NO\_REPORTED\_MISUNDERSTANDING & NRM \\
\bottomrule
\end{tabular}
\end{table}

\FloatBarrier
\begin{table*}[t]
\centering
\caption{Qualitative coding codebook using abbreviated labels.}
\label{tab:codebook_short}
\renewcommand{\arraystretch}{1.15}
\begin{tabular}{p{1.2cm}p{7.8cm}p{7.8cm}}
\toprule
\textbf{Code} & \textbf{Definition} & \textbf{Applied when participant indicated} \\
\midrule
PFA &
Product-level feedback on thesis, argument, organization,
language, or prompt engagement was perceived as accurate or useful. &
Praise of thesis, argumentation, grammar, coherence,
phrasing, or structure feedback.

\\

PAR &
The system correctly identified observable writing-process activity. &
Recognition of pauses, hesitation, backspacing,
revision frequency, or writing time.

\\

RRR &
The system correctly understood why a revision occurred. &
Revision for clarity, word choice, idea development,
restructuring, grammar, or tone.

\\

PPA &
The participant considered the process interpretation partly correct. &
Statements indicating partial agreement such as
``to some extent'' or ``partly.''

\\

PMI &
The system detected a process event but assigned an incorrect reason. &
Incorrect interpretation of pauses, deletion,
or revision intent.

\\

SRC &
Surface corrections were interpreted as deeper cognitive revision. &
Typos, spelling corrections, or typing errors treated as conceptual difficulty.

\\

TWC &
The system did not sufficiently consider writing constraints. &
Comments about rushing, limited time, or word limits.

\\

CEO &
Already corrected mistakes continued influencing evaluation. &
Participants noted that fixed errors were still criticized.

\\

ISM &
Feedback did not reflect rhetorical purpose or writing style. &
Misunderstanding of tone, examples, creativity, or intent.

\\

AGR &
Participants requested more specific improvement support. &
Requests for examples, replacements, highlighted errors,
or concrete suggestions.

\\

FRR &
Participants requested clearer feedback presentation. &
Requests for simpler wording or improved organization.

\\

RTR &
Participants requested clearer assessment criteria. &
Requests for rubrics, scoring criteria, or evaluation standards.

\\

SIR &
Participants requested technical/interface improvements. &
Requests for faster responses, better snapshots,
or interface improvements.

\\

NRM &
Participants explicitly reported no misunderstanding. &
Responses such as ``No'', ``N/A'', or no identified issue.

\\

\bottomrule
\end{tabular}
\end{table*}

\FloatBarrier
\section{Higher-Level Qualitative Themes}
\label{app:themes}
The 14 binary qualitative codes were organized into four
higher-level themes based on conceptual similarity and their role in
interpreting learner perceptions of process-aware feedback.
Table~\ref{tab:qual_themes} summarizes the relationship between the
higher-level themes and the corresponding abbreviated codes defined in
Appendix B.

\begin{table}[h]
\centering
\caption{Higher-level qualitative themes and associated codes.}
\label{tab:qual_themes}
\renewcommand{\arraystretch}{1.15}
\begin{tabular}{p{5.8cm}p{2.2cm}}
\toprule
\textbf{Higher-Level Theme} & \textbf{Associated Codes} \\
\midrule
Perceived Recognition of Product and Process Evidence & PFA, PAR, RRR, PPA \\[6pt]
Detection Without Consistently Correct Interpretation & PMI, SRC \\[6pt]
Contextual and Fairness Boundaries & TWC, CEO, ISM \\[6pt]
Requests for More Actionable and Usable Feedback & AGR, FRR, RTR, SIR \\[6pt]
\bottomrule
\end{tabular}
\end{table}

\FloatBarrier
The NRM code was retained as a descriptive indicator of explicit absence of perceived issues rather than a standalone thematic category.

\end{document}